\documentclass[preprint2]{aastex}
\usepackage{spr-astr-addons}
\usepackage{graphicx}
\newcommand{\gtapprox}{\raisebox{-0.5ex}{$\,\stackrel{>}{\scriptstyle
\sim}\,$}}
\newcommand{\ltapprox}{\raisebox{-0.5ex}{$\,\stackrel{<}{\scriptstyle
\sim}\,$}}

\begin{document}
\title{Extragalactic jets on subpc and large scales}

\shorttitle{Jets on different scales}        
\shortauthors{Tavecchio F.}

\author{F. Tavecchio} 
\affil{INAF/OAB, via E. Bianchi 46, 23807, Merate (Lc), Italy}
\email{fabrizio.tavecchio@brera.inaf.it}


x

\begin{abstract}
Jets can be probed in their innermost regions ($d \ltapprox 0.1$ pc)
through the study of the relativistically boosted emission of {\it
blazars}. On the other extreme of spatial scales, the study of
structure and dynamics of extragalactic relativistic jets received
renewed impulse after the discovery, made by {\it Chandra}, of bright
X-ray emission from regions at distances larger than hundreds of kpc
from the central engine. At both scales it is thus possible to infer
some of the basic parameters of the flow (speed, density, magnetic
field intensity, power). After a brief review of the available
observational evidence, I discuss how the comparison between the
physical quantities independently derived at the two scales can be
used to shed light on the global dynamics of the jet, from the
innermost regions to the hundreds of kpc scale.
\end{abstract}





\section{Introduction}
\label{intro}


Despite decades of intense study, the physical mechanisms acting
behind extragalactic relativistic jets are still poorly constrained
(e.g. De Young 2002). One of the major difficulties derives from the
fact that even the basic physical quantities characterizing the flow
(speed, density, composition, geometry and intensity of magnetic
fields) are still unaccessible to a direct measure. Classical studies
in the radio band, probing a small portion of the synchrotron emission
from relativistic electrons, are intrinsically limited and the
physical parameters of the plasma can be evaluated only with several
assumptions (e.g., equipartition between particles and magnetic
fields, minimum energy of the emitting electrons). Multifrequency
observations offer a good way to effectively overcome these
difficulties, helping us to disentangle the basic physical
quantities. In this respect, the best example is offered by the study
of jets in blazars, for which the modeling of the emission based on
the synchrotron and Inverse Compton mechanisms allows us to derive
robust estimates of the main parameters of the inner flow. In recent
years, the extension of the observations of large scale jets to the
optical and the X-ray band (possible thanks to the spatial resolution
of {\it Chandra}), have renewed the interest for the
field. Particularly exciting is the possibility that the X-ray
emission detected from jets in powerful quasars are produced by a
mechanims different from the synchrotron one, offering the possibility to
disentangle the basic physical parameters.


In the following I will discuss recent developments made in the study
of blazars (focusing in particular on the determination of the jet
power and the comparison with the accretion power) and in the
multifrequency investigation of large scale jets of powerful
quasars. Finally I will discuss how, comparing the physical properties
independently derived in the same jet in the blazar region and at
large scale, important clues on the global dynamics of the jet can be
derived.

\section{Blazars: probing the inner jet}

Blazars are excellent laboratories to study the innermost region of
jets. Their highly variable, relativistically boosted, non-thermal
continuum is produced by high-energy electrons (or pairs) in a
relativistic jet closely aligned with the line of sight (Blandford \&
Rees 1974). The small variability timescales, coupled with the
condition that the source is transparent to $\gamma$-rays, allows one
to constrain the distance of the emission region around $10^{17}$ cm
from the central Black Hole (Ghisellini \& Madau 1996), corresponding
to $10^2-10^3$ gravitational radii for typical BH masses. In the
widely discussed ``internal shock'' scenario (Ghisellini 1999, Spada
et al. 2001) this distance marks the region where shells of matter
ejected by the central engine with different velocities preferentially
collide (but see Katarzynski \& Ghisellini 2007), producing shocks at
which the magnetic field is amplified and relativistic electrons
responsible for the emission are accelerated. Alternatively, the
blazar emission could trace the instabilities in the flow arising at
the end of the acceleration region, where MHD processes accelerate the
flow to relativistic speeds (Sikora et al. 2005).

\subsection{Emission models - The ``blazar sequence''}

The Spectral Energy Distribution of blazars, covering all the
electromagnetic spectrum, from radio frequencies up to GeV-TeV
energies, is characterized by two broad bumps, the first peaking
between the IR and the X-ray band, the second one in the gamma-ray
domain. The first peak traces the synchrotron emission of relativistic
electrons, while the high-energy component is presumably due to the
Inverse Compton scattering between the same electron population and
soft photons, both the synchrotron photons themselves (SSC mechanism;
Maraschi, Ghisellini \& Celotti 1992) and ambient radiation entering
the jet (External Compton; Dermer \& Schlickeiser 1993, Sikora,
Begelman \& Rees 1994), although other possibilities, possibly
involving hadrons, cannot be ruled out (e.g., Mannheim 1993, Aharonian
2000, M{\"u}cke et al. 2003). In low power sources (generally lineless
BL Lac objects) it is assumed that the IC component is dominated by
the SSC process, while in powerful quasars (mostly Flat Spectrum Radio
Quasars), characterized by the presence of bright emission lines, the
EC process likely dominates. The simplest model (``one-zone'') assumes
that the bulk of the emission is produced within a single region.

\begin{figure}[ht!]
\begin{center}
\includegraphics[width=0.48\textwidth]{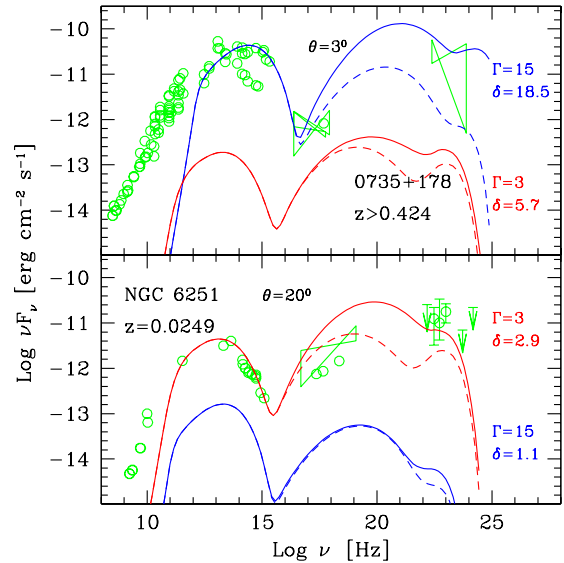}
\end{center}
\caption{{\it Upper panel:} Spectral Energy Distribution of the BL Lac
object 0735+178 (green points), together with the results of the
emission model for a structured jet (Ghisellini et al. 2005),
comprising the contribution of the fast ($\Gamma =15$) spine (blue)
and the slower ($\Gamma =3$) layer (red), assuming a small angle of
view ($\theta =3^o$) as typically derived for blazars. The dashed
lines report the Inverse Compton emission calculated only taking into
account the synchrotron photons produced locally. {\it Lower panel:}
the expected SED from the same jet observed at a larger angle of view
($\theta=20^o$). In this case the emission is dominated by the layer
(red), since the observer lies outside the narrow emission cone of the
stronger beamed spine (blue). For comparison, the green points report
the SED of the FR I jet in NGC 6251.}
\label{fig:6}       
\end{figure}

As shown by Fossati et al. (1998), the position of the peaks in the
SED is related to the luminosity of the emission (the so called
``blazar sequence'', see also Sambruna, these proceedings). Powerful
sources have both peaks located at low frequencies (IR and MeV
band). Moving from high to low luminosities, both peaks shift at
larger frequencies reaching, in the low luminousity BL Lac objects, the
X-ray band (synchrotron) and the TeV band (IC). A trend is also
present in the relative importance of the synchrotron and IC
peaks. The latter dominates the emission of FSQRs, while in low power
sources both components have almost the same importance (although for
some TeV sources there are indications that the high-energy component
can dominate during flares, e.g. Foschini et al. 2007).

It is worth noting that a trend between the peak of the emission and
the total luminosity has been derived also for Gamma Ray Burst
(Ghirlanda et al. 2004), for which the emission is believed to be
produced inside an ultra-relativistic (bulk Lorentz factors
$\Gamma=100-10^3$) jet produced during the collapse of massive
stars. In this case, however, the two quantities are correlated, the
most powerful sources showing the peak at high-energy and {\it
vice-versa}.

\subsection{Jet speed, magnetic fields, power and composition}

The description of the blazars SED with the synchrotron and IC
components allows us to derive with some confidence the values of the
basic physical quantities of the jet (e.g., Ghisellini et al. 1998,
Tavecchio et al. 2000a, 2001, Kubo et al. 1998, Sikora \& Madejski 2001). In
particular:

\noindent
$\bullet$ Bulk Lorentz factors in the interval $\Gamma =10-20$ are
commonly derived, consistent with those inferred at pc scale through
VLBI studies (e.g., Kellermann et al. 2004). However, larger values
(up to $\Gamma \gtapprox 50$) are required by synchro-SSC models of
some TeV BL Lac (Krawczynski et al. 2002, Konopelko et
al. 2003). These values are in contrast with the small, often
subluminal, apparent speeds measured with VLBI in these sources (e.g.,
Edwards \& Piner 2002, Piner \& Edwards 2004). A way to explain this
discrepancy is to admit that the jet decelerates between the subpc
scale and the VLBI scale (Georganopoulos \& Kazanas 2003). A {\it
bonus} of this interpretation is that if one takes into account in the
calculation of the IC emission the target photons coming from the
outer, decelerating, portion of the jet, it is possible to decrease
the required Lorentz factor to ``standard'' values $\Gamma \ltapprox
20$. Another possibilities along the same lines is to assume a
structured jet, with a fast spine surrounded by a slower layer
(Ghisellini et al. 2005). Apart from decrease the required Lorentz
factor, the strong radiative coupling between the fast spine and the
slow layer offers the possibility to justify the assumed deceleration
of the spine as effect of the radiation drag ({\it Compton drag}
effect). The presence of such a structure is also supported by the
direct radio imaging of some jets (Giroletti et al. 2004) and is
required by the unification of the spectral properties of BL Lac
objects and their parent low power (FR I) radiogalaxies (Chiaberge et
al. 2000). Ghisellini et al. (2005) also note that the radiative
coupling between the spine and the layer can boost the IC emission
from both regions, with the consequence that FR I radiogalaxies could
be strong $\gamma -$ray emitters. An example is reported in
Fig.(\ref{fig:6})

\begin{figure}[ht!]
\begin{center}
\includegraphics[width=0.48\textwidth]{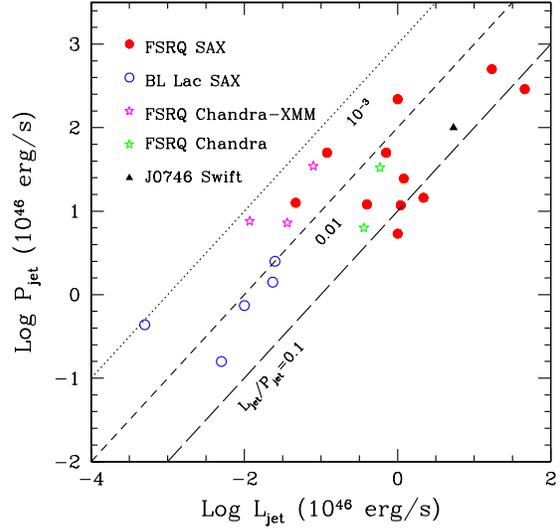}
\end{center}
\caption{The jet power $P_{\rm jet}$ compared with the
beaming-corrected radiative luminosity $L_{\rm jet}$ for a group of
blazars, including those originally discussed in Maraschi \& Tavecchio
(2003) plus other studied recently in Sambruna et al. (2006a,b) and
Tavecchio et al. (2007). Power is estimated assuming a composition of
1 proton per relativistic electron. The diagonal lines indicate
different values of the radiative efficiency ($L_{\rm jet}/P_{\rm
jet}$), from 10\% to 0.1\% (powers on both axis are expressed in units
of $10^{46}$ erg/s).}
\label{fig:1}       
\end{figure}

\noindent
$\bullet$ Magnetic fields with intensity ranging from $\sim 0.1$ G to
few G are usually derived. The magnetic field intensity increases with
the radiative luminosity of the source, being smaller in the low-power
BL Lacs and larger in the powerful FSRQs. In the latter sources the
values of the magnetic energy density are consistent with
equipartition with the energy density of relativistic electrons, while
in BL Lacs, especially the TeV emitting ones, values below
equipartition (by one order of magnitude) have been derived (Maraschi
et al. 1999, Kino et al. 2002).

\noindent
$\bullet$ The jet power can be inferred from the knowledge of the bulk
Lorentz factor and the particle density (e.g. Celotti \& Fabian
1993). In this respect a key information is the composition of the
jet. Unfortunately, our knowledge of the matter content of jets is
still rather poor. The observations only allow us to directly probe
the relativistic electrons (and the magnetic field) in the emitting
regions, but little is known on the possible existence of cold pairs
and protons and on the relative abundance of all the species. However,
some constraints on the matter composition can be derived using the
condition that the jet carries enough power to emit the radiation that
we observe. Using this condition, one can rule out (at least for
FSRQs) the possibility that the jet contains only the relativistic
electrons that we observe (e.g., Maraschi \& Tavecchio 2003, Sikora et
al. 2005). Therefore, another component, which carries most of the jet
power, is required. A direct possibility is to assume a ``normal''
plasma, with approximately 1 proton per relativistic electron.  In
Fig.(\ref{fig:1}) we report the comparison between the jet power
calculated assuming the e-p composition and the beaming-corrected
radiative luminosity inferred for a group of blazars, including the
sample of Maraschi \& Tavecchio (2003) and other sources recently
studied. FSRQs are mainly concentrated in the range of powers $P_{\rm
jet}=10^{47}-10^{48}$ erg/s, with few objects reaching larger powers,
while BL Lac jets have power of the order of $P_{\rm
jet}=10^{45}-10^{46}$ erg/s. The radiative efficiencies (defined as the
ratio $L_{\rm jet}/P_{\rm jet}$) range from at least 0.1\% to 10\%.

\begin{figure}[ht!]
\begin{center}
\includegraphics[width=0.48\textwidth]{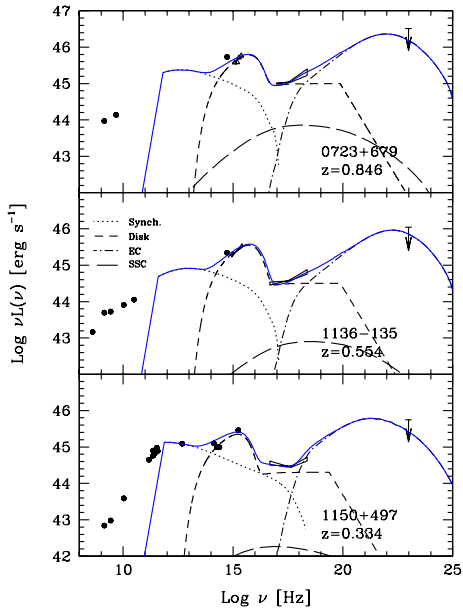}
\end{center}
\caption{Spectral Energy Distributions of the core of three radio-loud
quasars with detected X-ray emission from the large scale jet
(Sambruna et al. 2006a). The optical-UV data and the X-ray spectrum
are well modeled as a mix of disk and jet emission.  The lines report
the total emission model (solid) and the single components (dotted:
synchrotron; long dashed: SSC; short dashed: disk; dotted line: EC).}
\label{fig:5}       
\end{figure}

A possible alternative to protons is a dominant population of cold
(i.e. not relativistic) pairs. This possibility could be {\it
directly} tested, since a large amount of cold pairs would produce a
characteristic spectral bump in the soft X-ray band, through the
``bulk Compton'' process between the pairs and the soft ambient
photons (Begelman \& Sikora 1987, Celotti, Ghisellini \& Fabian
2007). Unfortunately, the detection of this spectral feature is
difficult, since it can be easily outshined by the IC emission from
the relativistic electrons.

A last possibility is that the jet is not matter-dominated, but that
the power is supported by dominant magnetic fields (Blandford 2002,
Lyutikov 2003). However, there are no strong observational evidences
supporting this possibility (Sikora et al. 2005).

\subsection{Jet power and accretion power}
\label{pjetpacc}

For the objects reported in Fig.(\ref{fig:1}) we also have information
on the luminosity of the accretion flow, either directly (when the
``blue-bump'' is visible) or indirectly through the measure of the
luminosity of the emission lines. As an example, in Fig.(\ref{fig:5})
we report the SEDs of three interesting sources recently studied by us
(Sambruna et al. 2006a), in which the presence of an important
contribution from the disk is clearly described by the shape of the
optical-UV continuum and supported by the presence of a weak iron line
in the X-ray spectrum, which can thus be interpreted as a mix of the
emission from the disk and the relativistic jet.

It is thus possible to compare the power carried away by the
relativistic flow with that supplied by the accreting
material. Previous studies indicate a correlation between the
accretion luminosity and the jet power estimated either through the
measure of the energy stored in radio-lobes (Rawlings \& Saunders
1991) or the modeling of the emission of the jet at VLBI scales
(Celotti, Padovani \& Ghisellini 1997). The existence of such
relations supports scenarios for jet production requiring a direct
link between outflows and accretion (e.g., Blandford \& Payne 1982).

\begin{figure}[ht!]
\begin{center}
\includegraphics[width=0.48\textwidth]{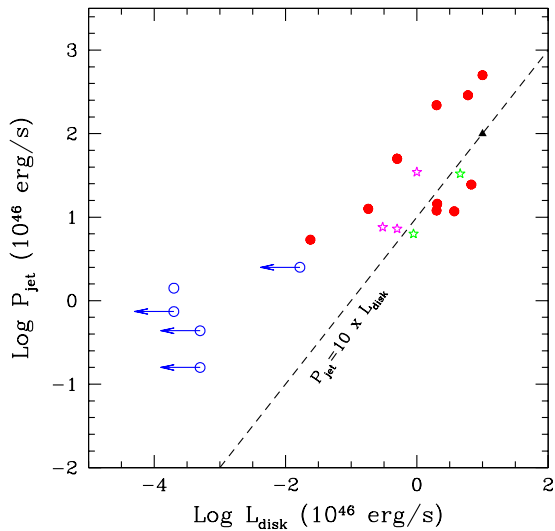}
\end{center}
\caption{Comparison between the jet power and the accretion
luminosity, for the same sources reported in Fig.(\ref{fig:1}). When
possible, the accretion luminosity has been directly inferred from the
``blue bump'' in the optical-UV band, otherwise we used the luminosity
of the broad emission lines, assuming a ratio of 0.1 between the
luminosity of the Broad Line Region and that of the accretion disk
(powers on both axis are expressed in units of $10^{46}$ erg/s).}
\label{fig:2}       
\end{figure}

The extension of this comparison to the blazar scale offers the
possibility of a more direct measure of the power, in regions closer
to the central BH. The comparison is reported in
Fig.(\ref{fig:2}). Most of the BL Lac objects do not have reliable
measures of line luminosities and only upper limits on the accretion
luminosity can be derived. A trend is clearly visible in
Fig.(\ref{fig:2}), with the accretion luminosity increasing with the
jet power.

Clearly the jets carries a sizable fraction of the accretion power. As
indicated by the dashed line, on average the jet power derived for
FSRQs is 10 times the accretion luminosity. For a standard accretion
efficiency of 10\%, this implies that in these systems, the jets
carries a power {\it of the same order} of the accretion power. For BL
Lac sources, apparently the jet carries a power much larger than that
associated to accretion flow. However, it is likely that the radiative
efficiency of the accretion flow in these low-power AGNs is smaller
than that of the standard accretion disk.

\begin{figure}[ht!]
\begin{center}
\includegraphics[width=0.48\textwidth]{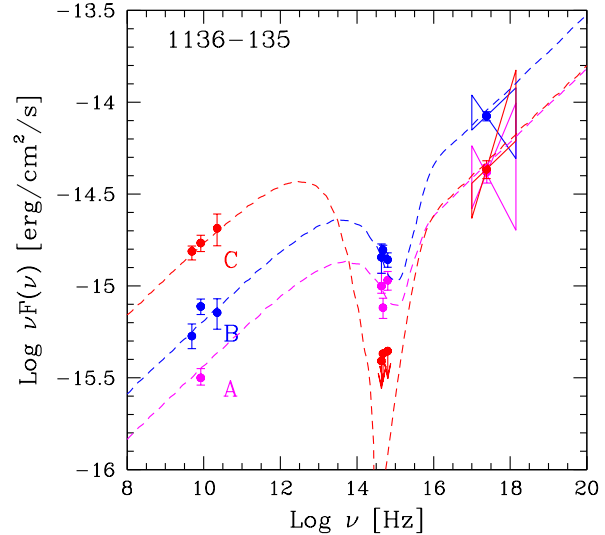}
\end{center}
\caption{SED of three knots of the jet associated to the quasar
1136-135, constructed with radio (VLA), optical ({\it HST}) and X-ray
({\it Chandra}) data. The optical data exclude a unique power-law
spectrum from radio to X-rays. The lines report the synchrotron and
IC/CMB emission which reproduce the data (from Sambruna et
al. 2006c).}
\label{fig:7}       
\end{figure}

\subsection{A unifying view}

The pieces of evidence collected above can be used to construct a
unifying scenario for the properties of the jet at small scales.

As pointed out by Ghisellini et al. (1998), the modeling of the SED of
a large group of sources suggests that the ``blazar sequence'' is
related to a trend in some of the physical parameters of the jet:
the energy of the electrons emitting at the peak of the SEDs
systematically increases from high to low power sources while, at the
same time, the energy density in the magnetic field and radiation
decreases. Ghisellini et al. (1998) argued that this trend can be the
result of the balance between the cooling rate (measured by the amount
of total energy density) and the (almost universal) acceleration rate
of the electrons.  The most powerful sources have a large amount of
magnetic and radiation energy density, determining a severe cooling
and thus a small value for the equilibrium Lorentz factor of the
electrons. On the contrary, BL Lacs are characterized by a low level
of cooling, explaining the large Lorentz factors of the electrons in
these sources. 

Along these lines, one can also envisage an evolutionary scenario in
which the progressive ``cleaning'' of the environment of the jet
during the cosmological evolution (e.g. Fabian et al. 1999) leads to a
decreasing accretion rate, which, in turn, implies a decrease of the
power of the jet, as suggested by the results of Sect.\ref{pjetpacc}.
The decreasing density of the AGN environment would also lead to a
minor photon density, which, as we discussed below, probably
regulates the cooling of the electrons, at least for powerful FSRQ, in
which cooling is dominated by the IC process.  The net result of this
process would be that an initially powerful FSRQ evolves into a
low-power, BL Lac object (Boettcher \& Dermer 2002, Cavaliere \&
D'Elia 2002).\\

Finally it is worth to stress that most of the discussion above is
based on modeling of SEDs often comprising non-simultaneous data. In
particular, $\gamma-$ray data are usually averages of the positive
detections, although it is well known that the high-energy emission is
highly variable.  {\it GLAST}, with its good sensitivity and the wide
field of view will allows us to better characterize the high-energy
component and to derive stronger constraints on the properties of the
jet at small scales.

\section{Large scale jets in quasars}

As discussed elsewhere in these proceedings (Marshall, Schwartz), the
detection of dozens of resolved jets in the X-rays, initiated a new
active field of research (a recent review is Harris \& Krawczynki
2006). While the multifrequency emission of low power (FR I) jets is
commonly interpreted as due to a unique (power-law or steepening
power-law) synchrotron component from the radio to the X-ray band
(e.g., Worrall et al. 2001), more debated is the interpretation of the
emission from large power (FR II) jets hosted by quasars. Clearly (see
Fig.\ref{fig:7}), a unique power-law component cannot reproduce the
multifrequency data, showing a well defined ``valley'' in the optical
region (Schwartz et al. 2000). The most direct explanation is that, in
analogy with blazars, we are observing two emission components from
the same electrons, the synchrotron emission accounting for the radio
and (in some cases) the optical emission and the IC mechanism
producing the bright X-ray component. The extreme power requirements
allow us to rule out SSC emission (Schwartz et al. 2000, Tavecchio et
al. 2004). An alternative is the IC scattering of photons of the
Cosmic Microwave Background. However, in order to reproduce the
luminous X-ray emission we have to assume that the CMB photons are
boosted in the jet frame, requiring relatively large bulk Lorentz
factors of the jet ($\Gamma \gtapprox 2-3$) at these large scales
(Tavecchio et al. 2000b, Celotti et al. 2001). If we further require
the equipartition between magnetic field and relativistic electrons
energy densities, we can completely determine the physical
parameters. In this framework the X-ray emission originates from
electrons belonging to the low-energy end of the energy distribution
(corresponding to Lorentz factors $\gamma \sim 10-20$), whose
synchrotron emission, being at very low frequencies, is unaccessible
to radio observations

In the following discussion we assume that the IC/CMB model is the
correct interpretation of the overall multifrequency emission of knots
in large scale jets of quasars. Criticisms and alternatives to this
interpretation can be found in Aharonian (2002), Stawarz et
al. (2004), Atoyan \& Dermer (2004), Kataoka \& Stawarz (2005).
Recent work specific for the jet of 3C~273 added further problematic
elements (Uchiyama et al. 2006, Jester et al. 2006). However, caution
should be used in generalizing these results obtained for a single
source, that for some aspects is an ``outlier'', to the entire
population of jets in quasars.

\subsection{Jet parameters}

The application of the IC/CMB model to the observed multifrequency
emission of a relatively large number of quasars provide quite
interesting constraints on the jet (see e.g., Sambruna et al. 2002,
2004, 2006c, Kataoka \& Stawarz 2005, Schwartz et al. 2006):

\begin{figure}[ht!]
\begin{center}
\includegraphics[width=0.48\textwidth]{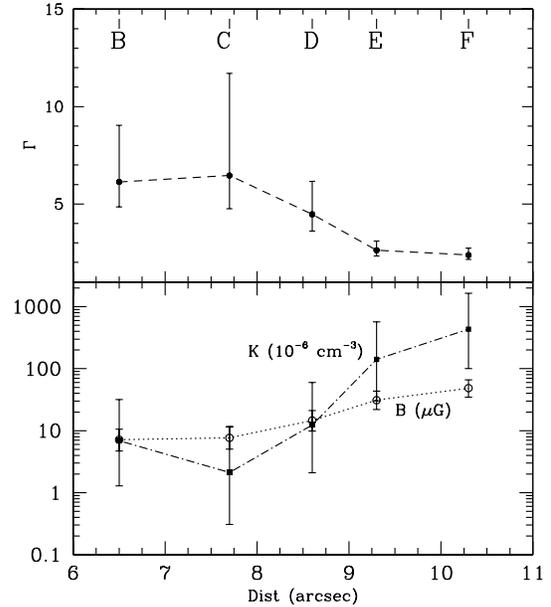}
\end{center}
\caption{Profiles of the Lorentz factor $\Gamma$ (top panel), magnetic
field intensity, $B$ and electron density, $K$ (lower panel) for
regions B--F of the jet of PKS 1136-135 estimated from the radiative
IC/CMB model (from Tavecchio et al. 2006). The decreasing Lorentz
factor marks the deceleration of the jet, accompanied by increasing of
both the magnetic field and electron density. The inferred
deceleration can be interpreted as due to the loading of the jet by
entrainment of external material.}
\label{fig:8}       
\end{figure}

\noindent
$\bullet $ The Lorentz factors obtained under the assumption of
equipartition generally lies in the range $\Gamma =5-15$. These values
are in several cases consistent with those at pc scale, required by
the observed superluminal speeds measured with VLBI. However, large
Lorentz factors seem to be in contrast with independent estimates
based on the jet to counter jet luminosity ratio of a sample of
radio-loud quasars, suggesting $\Gamma \ltapprox 3$ (Wardle \& Aaron
1997). A solution is to admit that the radio emission considered in
these estimates originates in a slower layer surrounding the faster
jet emitting at X-rays.

\begin{figure}[ht!]
\begin{center}
\includegraphics[width=0.48\textwidth]{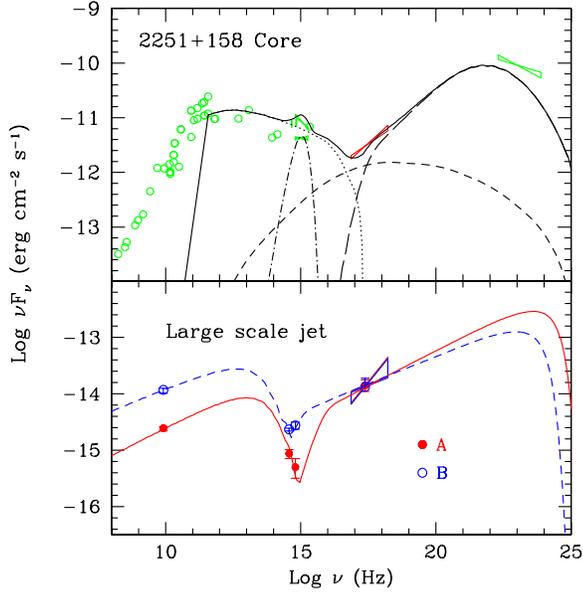}
\end{center}
\caption{Spectral Energy Distributions of different emission regions
for the sources 2251+158. {\it Upper panel}: {\it Chandra} X-ray
spectra of the blazar regions are shown together with non simultaneous
multifrequency data. The lines report the emission model used to
reproduce the data (solid) with the different components (dotted:
synchrotron; short dashed: SSC; long dashed: EC; dotted line: disk.)
{\it Lower panel}: radio, optical ({\it HST}) and X-ray ({\it
Chandra}) fluxes for different large scale jet knots. The latter show
the typical two component structure, well explained by the IC/CMB
model. Note that HST data, although mostly non-detections, provide
very important limits that prevent a single-component interpretation
of the SED.}
\label{fig:3}       
\end{figure}

In some cases there is a trend for $\Gamma $ to {\it decrease} along
the jet, indicating the possible deceleration of the flow
(Georganopoulos \& Kazanas 2004). The best studied case is that of
1136-135 (see Fig.\ref{fig:8}), for which we discussed the possibility
that the deceleration is induced by entrainment of external gas
(Tavecchio et al. 2006). However, this behaviour is not ubiquitous
(e.g., Schwartz, this volume).

\noindent
$\bullet $ The insensity of the magnetic field estimated assuming
equipartition generally lies in the range $B=10^{-6}-10^{-5}$ G. If
the equipartition condition is relaxed and if the Lorentz factor of
the jet is constrained to be $\Gamma \ltapprox 5$ as discussed above,
the resulting magnetic field is below equipartition, and the plasma is
strongy matter dominated (Kataoka \& Stawarz 2005). Sub-equipartition
fields are also suggested for the low-power jet in M87, based on the
current upper limit of its high-energy SSC emission (Stawarz et
al. 2005).  In general, relaxing the equipartition condition allowing
the electrons (the magnetic field) to dominate, implies a lower
(larger) bulk Lorentz factor and increases the jet power (see
Ghisellini \& Celotti 2001, Tavecchio et al. 2004).

In the jets for which there is evidence for deceleration, the inferred
magnetic field {\it increases} along the jet (Fig.\ref{fig:8}), as
expected from the adiabatic compression induced by the deceleration.

\noindent
$\bullet $ Rather interesting is the fact that the shape of optical
and X-ray continuum constrains the lower energy end of the electron
energy distribution (corresponding to Lorentz factors in the range
$\gamma _{\rm min}=5-20$), a quantity not easily accessible to the
direct measure with radio observations. The direct estimate of $\gamma
_{\rm min}$ allows us to robustly constrain the number of relativistic
electrons, particularly important in view of the determination of the
jet power.

\noindent
$\bullet $ The derived jet power (assuming the e-p composition) are
often rather large, in the range $P_{\rm jet}=10^{47}-10^{48}$ erg/s
(see also Ghisellini \& Celotti 2001). The large energetic requirement
is sometimes considered a problem for the IC/CMB interpretation (e.g.,
Atoyan \& Dermer 2004). However, all the sources for which the IC/CMB
model has been applied are powerful quasars and, moreover, these
values are consistent with the power derived for blazars
of comparable radio power (Fig.\ref{fig:1}; see also below).

\section{Jets from small to large scales}

Coupling information derived at subparsec scale and kiloparsec scale
for the same jet could have great potential to help in constructing a
global understanding of powerful extragalactic jets. This approach can
be fruitfully applied to those blazars showing a large-scale jet long
enough to be resolved by {\it Chandra}. Unfortunately, only few jets
can be studied on both scales, since the best studied blazars do not
tend to have well studied large-scale jets, precisely because the
former are the most closely aligned with the line of sight, reducing
the projected angular dimension of the large scale jet.

We first investigated (Tavecchio et al. 2004) two well known blazars
serendipitously belonging to the sample surveyed with {\it Chandra} by
Sambruna et al. (2004) and the study has been recently extended to
other 4 sources (Tavecchio et al. 2007). Similar results are reported
in Jorstad \& Marscher (2004, 2006). As an example we report in
Fig.(\ref{fig:3}) the SEDs of the blazar region and two knots of the
resolved jet of the quasar 2251+158, with the emission models used to
reproduce the data (from Tavecchio et al. 2007).

\begin{figure*}
\begin{center}
  \includegraphics[width=0.9\textwidth]{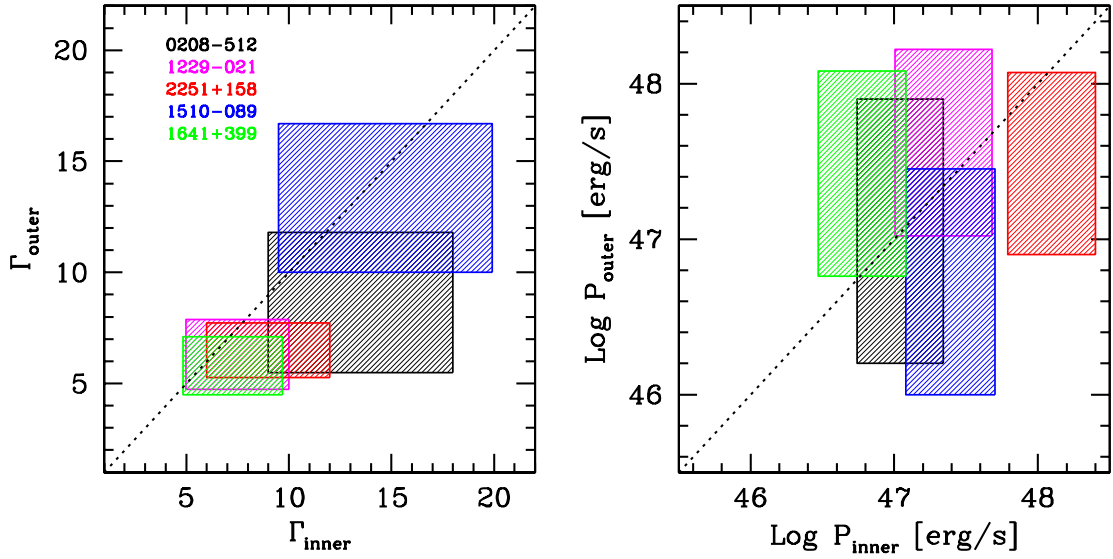}
\end{center}
\caption{Comparison between the jet Lorentz factor ({\it left panel})
and the jet power ({\it right panel }) evaluated independently for the
inner (blazar) jet and the kpc-scale jet for all sources for which
sufficient data exist. The rectangles include uncertainties due to
observational errors and to assumptions about the viewing angle, as
explained in the text.  The bulk Lorentz factor is consistent with
being constant from the subpc to the large scale scale jet and in any case
is still highly relativistic on the largest scale; the jet power is
also consistent with being constant up to very large scales, although
the uncertainties on the large scale estimates are rather large.}
\label{fig:4}       
\end{figure*}

From the independent modeling of the SED of the blazar and large scale
region we derived the basic parameters of the flow at the two scales
for the six sources. The comparison between Lorentz factors and powers
determined for the the blazar core ({\it inner}) and the large jet
knots ({\it outer}) is reported in Fig.(\ref{fig:4}), in which the
rectangles include the region of the plane allowed by the uncertainty
on the data and in the modeling. The plots show that, on average, the
Lorentz factor and the power derived at the two scales are in
agreement, suggesting that jets do not suffer important
deceleration and energy losses from the regions close to the black
hole to hundreds of kpc scale. However, the large uncertainty,
affecting in particular the derived power (in particular the values of
$P_{\rm outer }$, spanning in some cases a range larger than a
decade), prevent to draw a stronger conclusion.

\section{Conclusions: a simple view}

Our understanding of the physics associated to relativistic jets is
rapidly growing. Blazars allow us to investigate the innermost regions
of the jet, not far from the region where the flow is accelerated and
collimated (e.g., Junor et al. 1999). The modeling of the emission
observed from blazars provides important, albeit not conclusive, clues
on speed, power and composition of the flow and on the relationship
with the accretion feeding the central BH. On the other extreme of
spatial scales, multifrequency observations of large scale jets are
starting to shed some light on some of the basic problems.

The possibility to use the information collected at both scale can be
helpful in addressing some of the fundamental issues concerning
jet. Indeed, the results of the last section indicate that jets of
powerful quasars seem to evolve almost unperturbed from small to
large scales. However, as previously discussed, in some cases there is
complementary evidence suggesting that before its termination the jet
suffers important deceleration, marked by a decreasing Doppler factor
and the increase of magnetic field intensity and particle density
(Georganopoulos \& Kazanas 2004, Sambruna et al. 2006c). The
deceleration can be plausibly induced by entrainment of external gas
(Tavecchio et al. 2006), whose effects become important only when the
cumulative amount of entrained gas reaches some appreciable level
(Bicknell 1994). Moreover, the mixing layer thought to permit the
entrainment of the gas into the jet is believed to grow along the
jet. Therefore, entrainment can coexist with the evidence of the
conservation of power and speed, since the deceleration is expected to
become important only after some distance along the jet.

All these elements can be used to depict a simple scenario, in which
very powerful jets evolve freely, almost unperturbed, up to large
($\sim $100 kpc) scale, conserving the original power and speed (e.g.,
Blandford \& K\"{o}nigl 1979). In some cases (depending on external
conditions and jet power), the entrained mass becomes dynamically
important before the jet end, leading to the inferred deceleration
(e.g. the case of 1136-135, Tavecchio et al. 2006). It is tempting to
extend this view to include low-power FR I sources, characterized by a
small mass flux and therefore naturally more prone to deceleration.

Although model dependent, these result are quite interesting, and it
would be extremely important to confirm and strengthen them with
an enlarged sample of blazars with multifrequency observations of the
large scale jet. {\it GLAST} (scheduled to be launched at the end of
2007) is expected to greatly enlarge the number of known $\gamma$-ray
radio-loud AGNs. Likely, most of the sources associated to large scale
jets with known X-ray emission will be detected in the $\gamma$-ray
band, allowing to better characterize the SED of the core and
therefore increasing the number of objects suitable for this study.

\begin{acknowledgements}
I would like to thank L.. Maraschi, G. Ghisellini and R.M. Sambruna for
years of fruitful collaboration.  We acknowledge ASI for financial
support.
\end{acknowledgements}




\end{document}